\documentstyle[aaspp]{article}

\def\deg{\ifmmode^\circ\else$^\circ$\fi}

\def\kms{\ifmmode{\rm km\,s^{-1}}\else{km\,s$^{-1}$}\fi}
\def\r14{$r^{1/4}$}
\def\Ha{\ifmmode{\rm H\alpha}\else{H$\alpha$}\fi}
\def\HaNii{\ifmmode{\rm H\alpha + [N\,{\scriptsize II}]}\else{H$\alpha$ + [N\,{\scriptsize II}]}\fi}
\def\Nii{[N\,{\scriptsize II}]}

\lefthead{Balcells}
\righthead{Tails in NGC~3656}

\begin{document}

\title{Two tails in NGC~3656, and the major merger
origin of shell and minor axis dust lane ellipticals}

\author{Marc Balcells}
\affil{Instituto de Astrofisica de Canarias,
 C/ Via Lactea, S/N, 38200 La Laguna, Spain}

\begin{abstract}

I report on the discovery of two faint ($\sim$ 26.8 Rmag/arcsec$^2$)
tidal tails around the shell elliptical NGC~3656 (Arp~155).  
This galaxy had previously been interpreted as a case of accretion,
or minor merger.  
The two tidal tails are inconsistent with a minor merger, 
and point instead to a disk-disk major merger origin.  
NGC~3656 extends Toomre's merger sequence 
toward normal elliptical galaxies, 
and hints at a major merger origin for shells and minor-axis dust lanes.  
A dwarf galaxy lies at the tip of one of the tidal tails. 
A prominent shell, which shows sharp azymuthal color discontinuities,
belongs to a rotating dynamical component of young stars 
which includes the inner dust lane. 
\end{abstract}


\section{Introduction}
\label{Introduction}

Elliptical fine structure has been used to support the hypothesis
that ellipticals, whether normal or peculiar, lie along the 
disk-disk (D-D) merger evolutionary path 
(Schweizer, 1986; Schweizer \& Seitzer 1992).
But, when it comes to detailed modelling, 
minor mergers have been invoked more often than major mergers 
to account for shells (Quinn 1984, Dupraz \& Combes 1987), 
kinematically peculiar cores (Franx \& Illingworth 1988, 
Balcells \& Quinn 1990), 
and minor axis gaseous disks (Toomre 1977; Sage \& Galletta 1993).
The fact is, the presence of two opposing tidal tails
remains today the only reliable evidence for a D-D merger.  
And ellipticals, even peculiar ones, are not known to show tidal tails. 
As long as minor mergers provide plausible models for fine structure,
linking elliptical formation and major mergers remains 
more a matter of choice than one of necessity.

In this letter I report on the discovery of two tidal tails in 
the peculiar elliptical NGC~3656 (Balcells \& Stanford 1990, hereafter BS90).   
NGC~3656 has an elliptical body with a nearly edge-on minor axis dust lane, 
a system of shells, and a southern light condensation 
bound by a bright shell.
Balcells \& Sancisi (1996, hereafter BS96) report
the detection of $\sim 10^9 M_\odot$ of neutral hydrogen 
and of a similar amount of molecular hydrogen rotating along the minor axis.  
The HI has a lopsided distribution, and some material is
apparently associated with the prominent southern shell.
On the basis of the morphology, BS90 interpret this system as the result of 
the accretion of a smaller galaxy onto an already existing elliptical.  

The two tails reported here are incompatible with 
a minor merger, and point to a major merger origin for the galaxy and its 
other merger singatures, including the 
shells and the minor axis dust lane.  I provide evidence that the peculiar
colors in one shell are related to recent star formation
in a polar disk structure which includes the central dust lane,
and discuss the possible association of dwarfs to the western
tidal tail.  I compare NGC~3656 to prototype D-D
merger remnants and to NGC~5128.
A Hubble constant of 75 km\,s$^{-1}$\,Mpc$^{-1}$
and a distance of 40 Mpc (spatial scale 0.20 kpc/arcscec) are assumed.

\section{The data}
\label{TheData}

$R$-band imaging was carried out 
with the PF camera of the 2.5\,m Isaac Newton Telescope (INT)
at La Palma, in December 1995, using a (1024$\times$1024) Tektronix CCD
(scale 0.589 arcsec/pixel).  
Total exposure was 720 sec in $R_C$.
Flat-fielding was done using a combination of night sky and 
twilight sky exposures, and better than 0.2\% flatness was achieved.   
Seeing, as measured on stars in the frames, was 1.0 arcsec FWHM.
The photometric accuracy is 0.05 $R$ mag.  

A $B-R$ color-index image of NGC~3656 was derived from the $R$ image 
described above and a $B$ image taken with a (400$\times$590) EEV CCD
(scale 0.549 arcsec/pixel) at the INT PF in June 1990. 
Flat-fielding was good to $\sim 0.2$\%, and the weather 
was photometric.  The color zero-point uncertainty is $\sim$0.12 mag.

An \Ha\ image was obtained during the December 1995 run; 
two 900 sec exposures were taken 
through a 45 \AA\ wide, $\lambda$ = 6655 \AA\ 
filter which, in the F/3.29 beam of the INT PF at 0$^\circ$C temperature,
transmits with $\lambda_0 \approx 6633$ \AA\ (King et al. 1994).
This passband includes the \Ha\ and \Nii\ $\lambda$6583 lines of NGC~3656 (c$z$=2869 \kms, $\Delta v = 500$ \kms).  
The images were flat-fielded with twilight 
sky exposures, reaching 0.3\% accuracy over the entire field. 
I used the narrow-band and broad-band $R$ images to build
an \Ha\ line image and a red continuum image.  
Flux scaling was done by measuring fluxes in 24 stars in the frames.  
The uncertainty in the continuum scaling between $R$ and narrow band fluxes 
is the main source of error of the \Ha\ measured fluxes,
especially near zero levels of \Ha.  
Calibration was done by observing the star Feige 56 (Massey et al. 1988). 
For this mild starburst galaxy (\S\,\ref{StarFormation}), 
correction for \Nii\ is probably small.  
I took \Ha / (\HaNii) = 0.75 (Kennicutt 1983, 1992).
Foreground galactic extinction is negligible.


\section{The tidal tails}
\label{TidalTails}

\begin{figure}
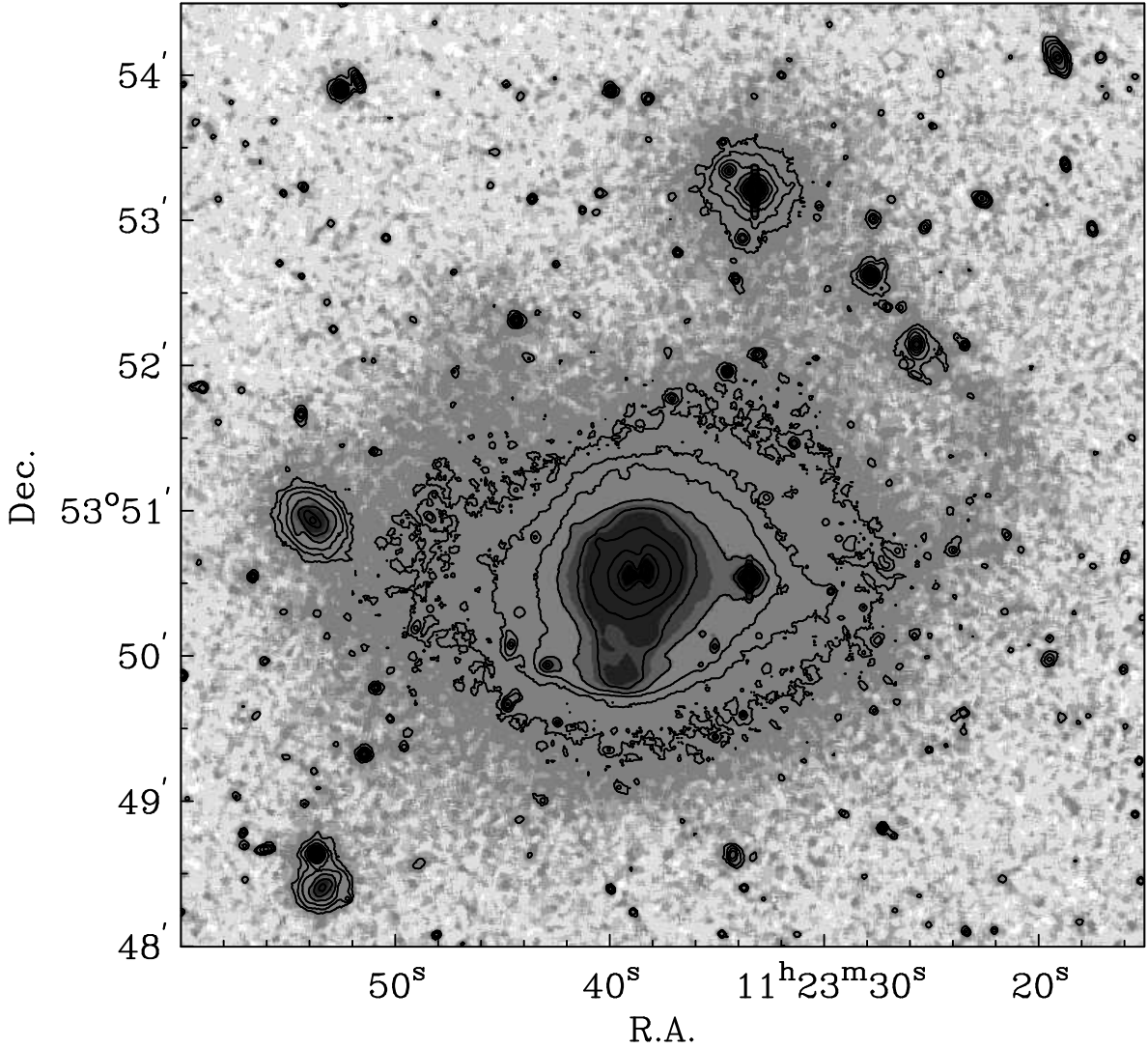

\caption{
	\label{Rband}
	Contours and gray-scale representation of an $R$-band image of
NGC~3656.  Lowest contour is 26.0 mag/arcsec$^2$,
contour spacing is 0.75 mag.  
Gray levels, drawn after applying a 5$\times$5 pixel median
filter,  correspond to 
$sky - 3\sigma$ ({\it white}), $sky - 2\sigma$, $sky - \sigma$,
$sky + \sigma, sky + 2\sigma, sky + 3\sigma$, to highlight the tails
against the sky background. 
Objects at $(\alpha, \delta)$ = 
(11:23:33, 53:50:33), (11:23:33.1, 53:53:13) and
(11:23:27.7,53:52:38) are stars. 
}
\end{figure}

Figure~\ref{Rband} (Plate 1) shows an $R$-band image of NGC~3656.
The central parts are contoured and the faint outer parts
are shown in grayscale.  

The first tidal tail emerges radially outward 
at a position angle PA = 260\deg\ and bends sharply to the NNE. 
It reaches the position of a dwarf galaxy at 
$(\alpha, \delta)$ = (11:23:25.6, 53:52:09; hereafter, Dw1). 
The tail might extend further up, but stellar images
make tracing the tail uncertain.  
No more traces of the tail exist beyond 
(11:23:34.4, 53:53:22), the position of a dwarf galaxy (hereafter Dw2) 
partially hidden in the stellar halo at (11:23:33.1, 53:53:13).

The mean surface brightness is 26.8 $R$mag/arcsec$^2$. 
The projected length is $\sim$ 182" = 36 kpc up to the first dwarf,
and it is $\sim 57$ kpc if the tail extends to the second dwarf.  
The tail has a mean width of only $\sim$15" = 3 kpc.  
From 15 kpc inward it broadens and brightens
(second contour in Figure~\ref{Rband}).  
After subtraction of an elliptical model of the galaxy, 
the tail can be traced inward to $\sim$9 kpc from the center.

This tail is fainter than its counterparts in {\it The Antennae} and
in NGC~7252, and, probably, intrinsically shorter despite 
the high geometric projection factor suggested by its sharp curvature. 
Nevertheless, the tail indicates that
one of the galaxies involved in the interaction 
was a disk galaxy which merged from a prograde orbit.  

The second tail can be recognized in the luminosity extension 
which protrudes to the NNE of the system.  
Although not as thin and well outlined as the west tail, 
it nevertheless bears the signature of a tail and not e.g. a shell, 
in that its outline is partially detached from that of the main body 
along the tail's west side as well as along its east side.  
It may be connected to the isophotal
distention to the south of the main body
(the prominent light condensation and shell) or to the distention to the east
traced by the faintest contours in Fig.~\ref{Rband}.
Its broader apparent width as compared to the west tail 
may be due to the viewing angle, as tidal tails have
flat,  ribbon-like shapes (Toomre \& Toomre 1972).
An intrinsically broader tail
could indicate a dynamically hotter precursor or weaker spin-orbit coupling.  
In either event, the pointiness of this tail is unlike the types of 
tidal distentions seen in interacting ellipticals
(see e.g. Borne et al. 1994).  In the latter,
isophotal twists caused by tidal distentions, which are sometimes 
quite pronounced, are gentle and gradual.  
In NGC~3656, the direction of maximum distention
moves from south (the southern shell) 
to east (lowest drawn contour in Fig.~\ref{Rband})
to NNE (outermost gray levels) for a total of over 135\deg.  

The simulations of Toomre \& Toomre (1972) already showed that 
two tails of similar lengths imply the merger of two disk galaxies.  
The two tails of NGC~3656 suggest therefore a D-D merger origin 
for the system.  
It is worth noting that the merger of a single spiral with an elliptical 
may also produce two tails if two pericenter passages take 
place before the merger is completed.  
However, in such a case the tails are highly dissimilar in 
length and surface brightness, due to the time elapsed between 
their formation; they are roughly coplanar, and do not point 
to each other.  None of these properties applies to the NGC~3656 tails.  

The shorter lengths in relation to the tails 
in prototype double-tail systems might indicate that
one of the galaxies was bulge-dominated, but might also indicate that 
the tails of NGC~3656 are older and, hence, more diluted.  

\section{The main body}
\label{MainBody}

\begin{figure}
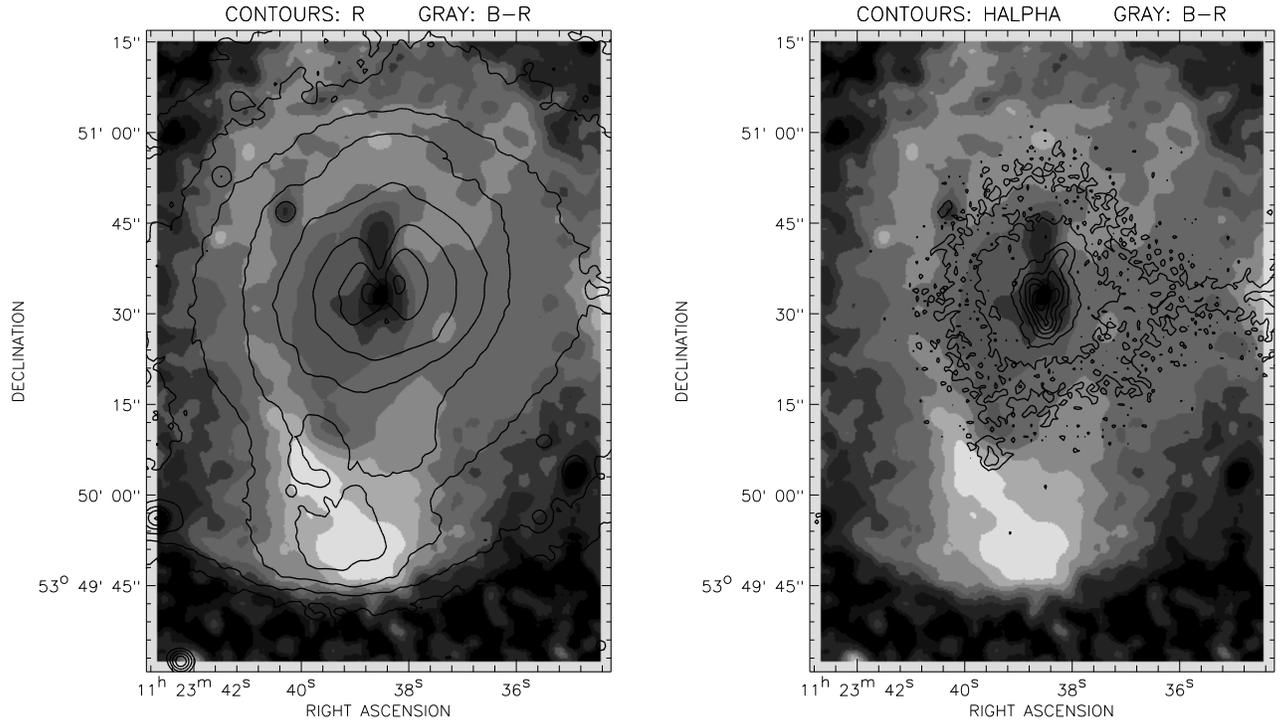

\caption{
	\label{MBoverlays}
	$(a)$ A gray-scale representation 
of the $B-R$ color map of NGC~3656, 
with $R$-band continuum contours overplotted.  
The color map was smoothed with a Gaussian kernel of $\sigma=3''$.
The lightest region to the South is $B-R \approx 0.9$.  
The central black region is $B-R \approx 2.1$.    
Colors to the South of the shell meaningless due to the 
low signal in the $B$ image.  
$(b)$ A contour representation of the H$\alpha$
emission, with gray scale representation of the $B-R$ color map. 
Gray levels as in $(a)$.  Contours are on an arbitrary
logarithmic scale. }
\end{figure}

We now focus on the main body, and search for additional 
clues about the origin of the system.  
Figure~\ref{MBoverlays} (Plate 2) shows a gray-scale display 
of the $B-R$ color index.
In Figure~\ref{MBoverlays}$a$ the contours trace the $R$-band continuum,
and in Figure~\ref{MBoverlays}$b$ the contours trace the \HaNii\ emission.

\begin{figure}
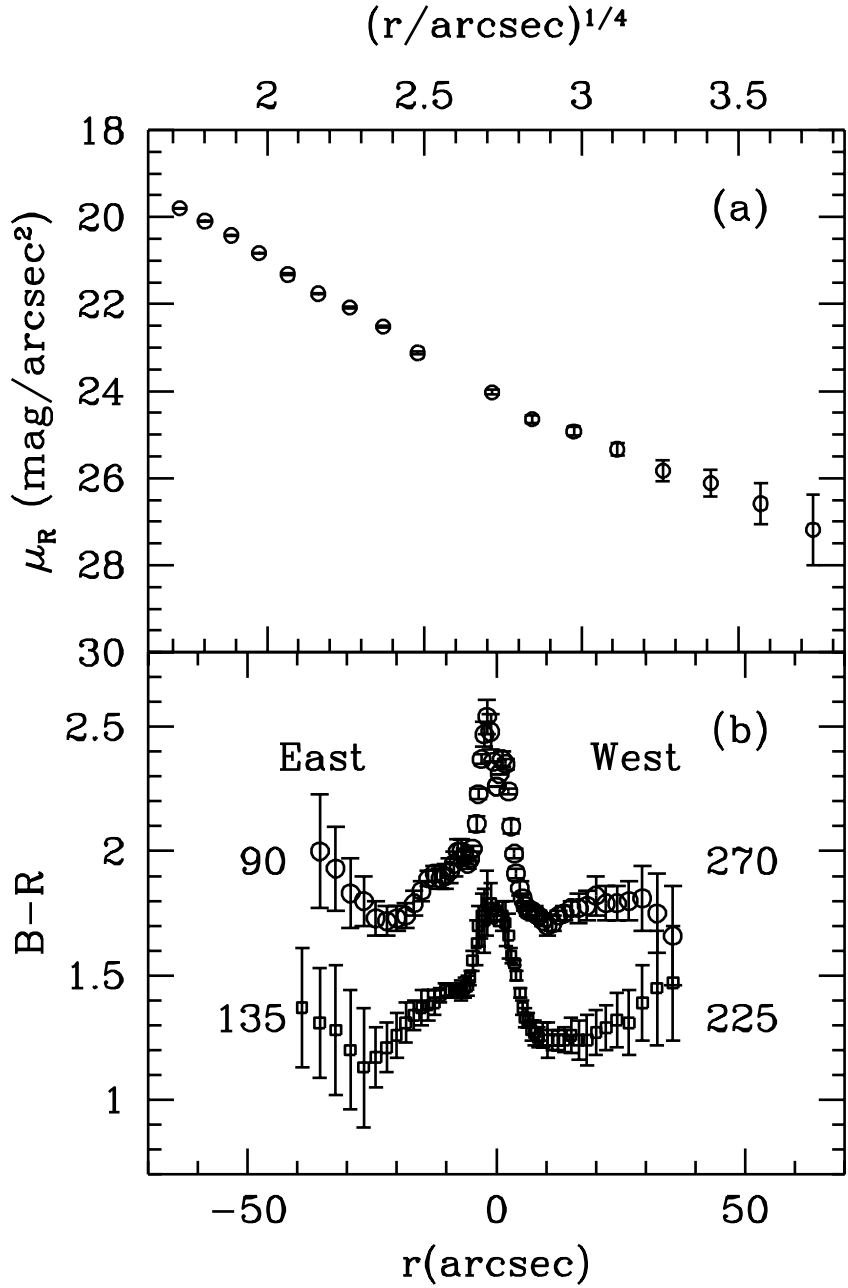

\caption{\label{Profiles}
$(a)$ $R$-band surface brightness profile derived from an
isophotal fit with centers, position angle and ellipticity fixed.  
$(b)$ Four $B-R$ color wedge-profiles along 45\deg\ wide wedge-shaped apertures 
with origin at the isophotal center of the galaxy (see text) and 
centered along the indicated position angles.
Negative abscissae denote eastern apertures.  The PA=$\pm 90\deg$ 
profiles have been shifted by +0.5 mag for clarity. 
}
\end{figure}

An isophotal fit to the galaxy image, after masking out the dust lane, 
the southern object and its shell, was carried out.  
The center,
position angle and ellipticity were left free first, then fixed
to values which minimized residuals in the main body:
$\epsilon = 0.20 \pm 0.04$, P.A. $= 110 \pm 5$ deg. 
Integrated magnitude out to 200 arcsec is $R = 11.81 \pm 0.05$. 
The best isophotal center lies in the middle of the dust lane, 
displaced
2.4 arcsec (0.47 kpc) north of the peak of the radio continuum emission 
(M\"ollenhoff, Hummel \& Bender 1992).
The main residuals occur at the southern luminosity condensation 
and along an arm which extends from this object toward the north 
on the east side of the galaxy (the "luminosity bridge" 
described by BS90).
The surface brightness profile (Figure~\ref{Profiles}$a$) 
closely follows an $r^{1/4}$ law out to $r = 60'' = 12$ kpc.
It was already known from $K$-band imaging that the stellar distribution 
in the inner parts follows an $r^{1/4}$ law (BS90).
The main interest in the present profile is that it shows that
the $r^{1/4}$ behaviour extends over the entire area of the optical shells. 
This rules out a face-on S0 classification for NGC~3656.  
The profile also shows that
the galaxy is much larger than previously suspected.  
The light distribution extends out to $r = 200'' = 40$ kpc,  
3 times farther away, and 3 mag deeper 
than e.g. the region imaged in Arp's (1966) atlas.  
In this outer region the profile shows a strong excess with respect 
to the $r^{1/4}$ law.
This region contains a significant 30\% of the total system light. 
The extended light distribution reaches a small barred spiral 
to the East (Fig.\,\ref{Rband}).  
Given that this galaxy is only mildly distorted,
and that it does not contain HI (BS96), it is unlikely that it
is responsible for the presence of shells, gas and dust in NGC~3656.


The $r^{1/4}$ profile suggests that material in the inner 60 arcsec 
has violently relaxed.  Probably,
the outer material did not take part in the violent relaxation 
because it was far from the center during the violent relaxation phase
(Binney \& Tremaine 1987).
In the main body, violent relaxation appears to be finished. 
The light distribution is smooth, 
the main isophotal irregularities being related to shells 
and to the subsystem defined by the dust lane 
and the southern luminosity condensation.  

The distribution of the $B-R$ color index is displayed in 
Figure~\ref{MBoverlays}.  
Colors in the main body are in the range $1.2\leq B-R\leq 1.4$,
typical of blue bulges (Peletier \& Balcells 1997), and
$\sim 0.3$ mag bluer than the colors of 
ellipticals of the same luminosity (Peletier et al. 1990).
The southern light condensation is distinctly bluer: $B-R \approx 0.9$, 
ie. comparable to the integrated colors 
of late-type spirals and irregulars (Buta \& Williams, 1995).

A ring with bluer colors extends from 
the southern light condensation through the east all the way to the north.
It has the same morphology 
as a region of strong positive residuals in the isophotal fit 
to the galaxy, and hence traces a distinct photometric component.
Figure~\ref{Profiles}$b$ shows four radial wedge profiles of the color map
at position angles 90\deg, 135\deg, 225\deg\ and 270\deg.   
Local minima occur at 20-30 arcsec on the eastern, and at 10--20 arcsec  
on the western apertures. 
The ring, hereafter referred to as "the blue ring", might
close on itself on the west side of the center, and it is offcentered. 
In view of the alignment of the blue ring and the inner dust lane,
we are inclined to believe that ring and dust lane are part of a single,
still forming dynamical system of gas (BS96) and young stars 
with red inner colors due to dust.  
The southern light condensation defines the edge of this disk.  

The distinct blue color of the southern light condensation
results in sharp color discontinuities in the shell which bounds
this object to the south see Fig.~\ref{MBoverlays}$a$). 
The alignment could be fortuitous, but it does suggest
an association between the entire shell and the blue ring.  
Such association points at a space-wrapped shell 
rather than a phase-wrapped shell,
and to recent localized star formation as responsible for
the azimuthal color variations.  
The evidence for star formation in the NGC~3656 shell
may provide an explanation for azimuthal color variations 
reported 
in shells of NGC~474 (Prieur 1988) and NGC~2865 (Fort et al. 1986).  
 
\section{Star formation}
\label{StarFormation}

The color distribution seen in the previous section 
suggests that star formation occurs within an organized 
dynamical subsystem in the merger remnant.  
We now map the star formation directly via H$\alpha$ maps.  

The net \HaNii\ emission is shown in Figure~\ref{MBoverlays}$b$.
It is strongly centrally concentrated; it peaks near the 
center of the dust lane on its eastern side, 
2.5 arcsec east of the peak of the radio continuum emission 
(M\"ollenhoff, Hummel \& Bender 1992)
and extends to the surrounding area.  
The peculiar geometry north of the center is real. 
It does not change under moderate variations of the 
alignment, the scaling or the seeing matching 
between line and continuum images.  
Dust extinction is surely responsible for this pattern,
as well as for the shift between the \Ha\ and the 
radio continuum peaks.
Estimating how much \HaNii\ flux we are missing is 
difficult because we don't know the distribution 
of the \Ha\ emitters within the dust lane, and because 
the shape and inclination of the dust lane is uncertain.  
Star formation probably occurs all along the dust lane,
judging from the similarities of the dust distribution 
and the radio continuum emission.

Low-level \Ha\ emission is seen outside the dust lane.  
Checks were carried out
to verify that this emission does not result  from incorrect
broad- to narrow-band scaling.
The extended emission becomes fainter outward, and
may extend into the blue ring, but at a level below
our detection.  
At the southern light condensation 
the net flux is also consistent with zero.  

The measured \HaNii\ flux in erg/s/cm$^2$ is 
$\log f$(\HaNii) = -12.74, 
and the implied luminosity, in erg/s, 
is $\log L$(\HaNii) = 40.51.  
The measured \Ha\ Luminosity is one to two orders of magnitude
above the \Ha\ luminosities of normal ellipticals  
(Goudfrooij et al. 1994), and it is within the range of 
\Ha\ luminosities of normal and interacting spiral galaxies
(Kennicutt et al. 1987, Bushouse 1987).  
It is similar to the \Ha\ luminosity of the 
prototype merger remnant NGC~3921 (Schweizer 1996). 
The implied (lower limit) total star formation rate (SFR) 
is of order 0.05 M$_\odot$\,yr$^{-1}$
using the scaling from Kennicutt (1983), and
0.14 M$_\odot$\,yr$^{-1}$ using the scaling from Bushouse (1987).

\section{The dwarfs}
\label{TheDwarfs}

It is intriguing that one, possibly two dwarf galaxies lie along 
the western tidal tail.  The alignment could be fortuitous, 
given the moderate density of non-stellar sources in the field 
(Deeg et al., 1997). 
There is compelling evidence that dwarf galaxies
may form in tidal tails (Mirabel et al. 1991).
Dw1 ($R$ = 17.2 mag) has an exponential 
surface brightness profile between 4" and 10",
with extrapolated $\mu_{0,R} = 21.0$, and a central cusp.
Dw2 has an $r{1/4}$ profile, hence it could be a background elliptical. 

\section{Discussion}
\label{Discussion}

The newly reported tidal tails are welcome additions
to the landscape of merger signatures of NGC~3656.  
By pointing to a major merger,
the tails break the ambiguities regarding the type of merger 
which generated the peculiarites seen in NGC~3656.
Because of the tails, NGC~3656 lies
beyond the end-point of Toomre's (1977) D-D merger sequence,
effectively extending the D-D merger sequence closer
to bona-fide, albeit peculiar, ellipticals with shells 
and gaseous disks.  
We may then compare the properties of NGC~3656 
to those of recent D-D remnants such as NGC~7252 and NGC~3921, 
and to those of shell and dust lane ellipticals such as NGC~5128, 
and seek evolutionary links between them.  

Kinematic data, perhaps from HI, and detailed merger models 
will give important checks on the major merger origin of NGC~3656.  
Preliminary $N$-body modelling indicates that the tail geometry
is easy to reproduce with a D-D merger, but
matching the internal kinematics 
will surely prove more challenging.  $N$-body modelling may 
tell us whether the apparent short length of the tails reflects
a large projection factor, an early-type disk progenitor, 
a weak spin-orbit coupling, or simple aging of the tails.  

If NGC~3656 formed in a major merger, then it follows that the remainder of 
the merger signatures in the galaxy
should have originated in the same D-D merger. 
This applies to the shell system, the minor axis dusty disk,
the star formation activity, and the kinematically decoupled core.
The latter two are expected results of major mergers 
(e.g. Hernquist \& Barnes 1991). 
But shells and minor-axis dusty disks 
have long been associated to "conservative" minor mergers 
which build a peculiar subsystem
causing little transformation of the already existing elliptical
(e.g. Quinn 1984; Bertola 1990).  
NGC~3656 suggests that Nature 
is capable of generating fine structure
in the course of the more catastrophic major mergers. 
The key to the formation of coherent dynamical substructures
during major mergers could be related to the
return of tidal material onto a partially relaxed main body
(Hibbard \& Mihos 1995).
That returning tidal material is both coherent 
and spread over orbital phase may help
the formation of minor axis disks.  

NGC~3656 has several similarities to the prototype young merger
remnants NGC~7252 and NGC~3921.  From the smoothness of the light 
distributions, NGC~3656 is the more evolved of the three.
Balmer absorption is seen in the stellar spectra of the three galaxies.
Like NGC~7252, NGC~3656 has a disk of ionized gas.
In the former, the disk has a central minimum, whereas,
in the latter, \Ha\ peaks at the center.  A central peak
is seen in the ionized gas in NGC~3921, 
where the gas does not form a disk (Schweizer 1996). 
Modelling will clarify if the central density differences are due 
to one of the merging galaxies being bulge-dominated.  
Interestingly, new VLA D data 
(Balcells, van Gorkom \& Sancisi, in preparation)
shows that, as is the case in NGC~3921, one tidal tail 
in NGC~3656 is gas rich whereas the other is devoid of neutral gas,
pointing to an Sa-Sc or an S0-Sc merger. 
If we take NGC~3656 as a model, then NGC~3921 will develop
an organized gas disk as its tails partly fall onto the galaxy
and partly escape.  

The tails in NGC~3656 lend credence to  
major merger scenarios of the origin of shells and minor axis gaseous disks
in normal ellipticals.  
The well studied case of NGC~5128 (Centaurus~A) 
deserves special mention. 
With two tidal tails and a lopsided minor axis disk,
the merger record in NGC~3656
is more prominent than that in NGC~5128.  But otherwise, 
the similarities  between NGC~3656 and NGC~5128 are striking.
Both galaxies have a warped minor axis dusty gaseous disk
which emits in \Ha, both have optical shells, 
both have similar stellar and \Ha\ luminosities, 
similar effective radii (Dufour et al. 1979)
similar velocity dispersions (BS90; Wilkinson et al. 1986)
and similar amounts of HI, 
some of which appears to be associated with optical shells
(BS96; Schiminovich et al. 1994);  
their ellipticity profiles are similar, tracing 
a nearly round inner body which becomes flatter outward.
The only major differences are the absence 
of a double lobe radio source in NGC~3656, 
the absence of tidal tails in NGC~5128, 
and the more lobsided distribution 
of the minor axis disk in NGC~3656.  
The similarities give credence to the hypothesis that, like NGC~3656, 
Cen~A formed in a major D-D merger: 
a truly new arrival in the neighbourhood of the Local Group.
Sparke (1996) has shown that, in NGC~5128, 
a warped gaseous disk precessing in the potential of 
the galaxy, assumed oblate, reproduces the distribution 
of both the ionized gas along the minor axis and the outer
HI associated with the major axis shells, 
once the varying ellipticity of the galaxy is taken into account. 
Outer shells and inner disk may thus be part
of a single dynamical system which warps by over 90\deg. 
That such gas traces returning tidal material 
after a major merger has been advocated several times
(Schweizer 1995; van Gorkom \& Schiminovich 1996).  
The discovery of tidal tails in NGC~3656, a galaxy in so many ways
similar to NGC~5128, provides additional clues that
the major merger hypothesis is on the right track.  

\acknowledgements
I thank Jacqueline van Gorkom, Renzo Sancisi, 
Reynier Peletier, Rob Swaters and Emmanuel Vassiliadis
for support and encouragement in many ways.
The isophotal fits were performed with 
GALPHOT (J\"orgensen et al. 1992).

\newpage

\eject


\begin{references}
\reference{}Arp, H. C. 1966, ApJS, 14, 1
\reference{}Balcells, M., Peletier, R. F. 1994, AJ, 107, 135
\reference{}Balcells, M., Quinn, P. J. 1990, ApJ, 361, 381
\reference{}Balcells, M., Sancisi, R. 1996, AJ, 111, 1053 (BS96)
\reference{}Balcells, M., Stanford, S. A. 1990, ApJ, 362, 443 (BS90)
\reference{}Barnes, J. E. 1988, ApJ, 331, 699
\reference{}Bertola, F., 1990, in {\it Dynamics and interactions of galaxies},
ed. R. Wielen, Heidelberg:Springer-Verlag, 249
\reference{}Borne, K. D., Balcells, M., Hoessel, J. G., McMaster, 
1994, ApJ, 435, 79
\reference{}Bushouse, H. A. 1987, ApJ, 320, 49
\reference{}Buta, R., Williams, K. L. 1995, AJ, 109, 543
\reference{}Deeg, H. Mu\~noz-Tu\~n\'on, C., Tenorio-Tagle, G., 
Telles, E., Vilchez, J.M., Rodriguez-Espinosa,
J.M., Duc., P.A. and Mirabel, I.F. 1997, A\&A, submitted
\reference{}Dufour, R. J., van den Bergh, S., Harvel, C. A., 
Martins, D. H., Schiffer, F. H., Talbot, R. J., Talent, D. J., Wells, D. C. 
1979, AJ, 84, 281
\reference{}Dupraz, C., Combes, F. 1987, A\&A, 185, L1
\reference{}Fort, B., Prieur, J. L., Carter, D., Meatheringham, S. J., 
Vigroux, L, 1986, ApJ, 306, 110
\reference{}Franx, M., Illingworth, G. I. 1988, ApJ, 327, L55. 
\reference{}Goudfrooij, P., Hansen, L., J\o rgensen, H. E., N\o rgaard-Nielsen, H. U. 
1994, A\&AS, 105, 341
\reference{}Hernquist, L., Barnes, J. E. 1991, Nat, 354, 210
\reference{}Hernquist, L., Quinn, P. J. 1987, ApJ, 312, 1
\reference{}Hernquist, L., Spergel, D. N. 1992, ApJ, 399, L117
\reference{}Hibbard, J. E., Mihos, J. C. 1995, AJ, 110, 140
\reference{}J\"orgensen, I., Franx, M., Kjaergaard, P. 1992, A\&AS, 95, 489
\reference{}Kennicutt, R. C. 1983, ApJ, 272, 54
\reference{}Kennicutt, R. C. 1992, ApJS, 79, 255
\reference{}Kennicutt, R. C., Keel, W. C., van der Hulst, J. M., Hummel, E., 
Roettiger, K. A. 1987, AJ, 93, 1011
\reference{}King, D. L., Rand, R. J., Balcells, M. 1994, Spectrum, 2, 20
\reference{}Massey, P., Strobel, K., Barnes, J. V., Anderson, E. 1988, ApJ, 328, 315
\reference{}Mirabel, I. F., Lutz, D., Maza, J. 1991, A\&A, 243, 367
\reference{}M\"ollenhoff, C., Hummel, E., Bender, R. 1992, A\&A, 255, 35
\reference{}Peletier, R. F., Balcells, M. 1997, New Ast., 1(4), 349
\reference{}Peletier, R. F., Davies, R. L., Illingworth, G., Davies, L., Cawson, M. 1990, AJ, 100, 1091
\reference{} Prieur, J. L. 1988, Thesis, Universite Paul Sabatier 
(Toulouse).
\reference{}Quinn, P. J. 1984, ApJ, 279, 596
\reference{}Sage, L. J., Galletta, G. 1993, ApJ, 419, 544
\reference{}Schiminovich, D., van Gorkom, J. H., van der Hulst, J. M., 
Kasow, S. 1994, ApJ, 423, L101
\reference{}Schweizer, F. 1986, Sci, 231, 227
\reference{}Schweizer, F. 1990, in {\it Dynamics and interactions of galaxies},
ed. R. Wielen (Heidelberg:Springer-Verlag), 270
\reference{}Schweizer, F. 1995, in {\it Stellar populations}, 
IAU Symp. 164, eds. P. van der Kruit \& G. Gilmore, 
(Kluwer: Dordrecht), 275
\reference{}Schweizer, F. 1996, AJ, 111, 109
\reference{}Schweizer, F., Seitzer, P. 1992, AJ, 104, 1039
\reference{}Sparke, L. S. 1996, ApJ, 473, 810
\reference{}Toomre, A. 1977, in {\it Evolution of Galaxies and stellar populations},
ed. xxx
\reference{}Toomre, A., Toomre, J. 1972, ApJ, 178, 623
\reference{}Van Gorkom, J. H., Schiminovich, D. 1997, 
in {\it The Nature of Elliptical Galaxies}, 2nd. Stromlo Symposium, 
eds. M. Arnaboldi, G. S. da Costa \& P. Saha, in press
\reference{}Wilkinson, A., Sharples, R. M., Fosbury, R. A. E., Wallace, P. T. 
1986, MNRAS, 218, 297
\end{references}
\end{document}